\documentclass[aps,pra,twocolumn,showpacs,superscriptaddress,eqsecnum]{revtex4}
\usepackage{epsfig}
\usepackage{graphicx}
\usepackage{bm}
\begin{document}
\date{\today}
\title{Universal discriminator for completely unknown optical qubits}

\author {Bing He} 
\email {bhe98@earthlink.net}
\affiliation{Department of Physics and Astronomy, Hunter College of the City University of New York, 
 695 Park Avenue, New York, NY 10021}
 
\author{J\'{a}nos A. Bergou}
\affiliation{Department of Physics and Astronomy, Hunter College of the City University of New York, 695 Park Avenue, New York, NY 10021}
 
\author{Yuhang Ren}
\affiliation{Department of Physics and Astronomy, Hunter College of the City University of New York, 695 Park Avenue, New York, NY 10021}

\begin{abstract}
We propose an experimental setup that is capable of unambiguously discriminating any pair of linearly independent single photon polarization qubits,
about which we don't have any knowledge except that an extra pair of these unknown states are provided as the reference. This setup, which is constructed 
with optical CNOT gates, weak cross Kerr non-linearities, Bell state analysers and other linear optical elements, transforms the unknown triple photon 
input states to the corresponding single photon states to be deterministically processed by linear optics circuit. 
The optimal discrimination of the unknown states is achieved by this setup. 
\end{abstract}

\pacs{03.67.Hk, 03.67.Lx}

\date{\today}
\maketitle

\section{Introduction} \label{section1}
When we discriminate a pair of completely unknown quantum states $|\psi_i\rangle$, for $i=1,2$, the only possible way is to have copies of these unknown 
states as the reference, and then we can design the proper measurement that is capable of distinguishing between the inputs, e.g., $|\psi_1\rangle|\psi_2\rangle|\psi_i\rangle$, in which the data states $|\psi_i\rangle$ are appended to the reference copies to form the 
states with the permutation symmetry. If we require that the incoming unknown data states should be unambiguously discriminated, our device will output 
three possible results: 1 corresponding to the input being $|\psi_1\rangle$, 2 to that being $|\psi_2\rangle$ , and 0 for inconclusiveness. 
The optimal unambiguous measurement of these symmetric input states has been studied in both the Bayesian \cite{b-h05} and the minimax approaches \cite{zy}.

Another feature we need to have for such a device is its universality: it will perform optimally for any pair of unknown states. Since the input 
states are unknown, the only quantity we can use to indicate how well the device works is the average success probability over the Hilbert spaces of 
all randomly distributed $|\psi_1\rangle$ and $|\psi_2\rangle$, which are discriminated by this single device. For the discrimination of two unknown
qubits with one copy of them as the reference, the optimal average succes probaility can be as large as $1/6$, if these two unknown qubits appear with
an equal {\it a priori} probability \cite{b-h05}.

There are ways to improve this optimal average success probability. We can prepare multiple copies of reference and apply different types of measurement, 
i.e., projection value measure (PVM) and positive operator value measure (POVM) working in different range of {\it a priori} probability of the inputs, to significantly 
increase this value, if we follow the Bayesian approach to deal with the problem  \cite{bhe06-1}. For the discriminators to unambiguously identify the 
averaged unknown input states, the IDP (Ivanovic-Dieks-Peres) upper bound \cite{ivanovic,dieks,peres} of success probability in discriminating a pair of known states can be 
achieved if the copy number of the reference states goes to infinity \cite {hayashi}. To increase the optimal average success probability further, we can
use both multiple reference and data copies, so that the device will realize the deterministic discrimiantion of any pair of unknown states in the limit
of infinite reference and data copy numbers together \cite{bhe07-1}.

This type of unknown state discriminators can be of potential value in quantum communication and quantum computing, so it is interesting for us to study 
their feasible physical implementations. In \cite {bhe06-2} we propose a pure optical way based on interferometry \cite{reck} to deterministically realize the optimal measurements in \cite{b-h05} under the assumption that all the symmetric input states can be prepared as single photon states in multi-rail representation. The quantum circuit to realize the unknown qubits discrimination without 
achieving the optimal measurement and the possible implementation of its ingredient logic gates in ion traps are studied in \cite {j-o}. 
The design of the unitary operators to implement the discrimination of a pair of unknown qubits is also discussed in \cite{p-p}.

In this paper, we design an experimental setup to unambiguously and optimally discriminate the symmetric input states 
$|\psi_1\rangle|\psi_2\rangle|\psi_i\rangle$ ($i=1,2$) prepared with {\it any} pair of linearly independent unknown single photon polarization qubits. 
This is a device to process the unknown triple photon input signals, and all the technologies involved have been experimentally realized thus far. 
After the transformation from the triple photon inputs to single photon signals, only the ancilla of vacum state will be used in the realization of the optimal POVM the device performs.  
The rest of the paper is organized as follows: in Sec. II we present the physical basis to realize the optimal measurements designed to discriminate the general unknown qubits, which we intend to realize for the polarization photon systems; in Sec. III we look at the state preparation part of our device, where the triple photon input
states are transformed to the single photon states through a combined teleportation protocol, so that they can be deterministically processed by multi-port interferometer; 
the simplest linear optics circuit to finally determine the incoming unknown date qubit is given in Sec. IV; and finally we give some conclusive discussions in the last section.

\section{Physical Basis to Discriminate Unknown Optical Qubits} \label{section2}

We first give a brief review of the optimal measurements for discriminating any pair of unknown states we aim to realize for single 
photon polarization qubits. The input signals processed by our device are given as follows:
\begin{eqnarray}
|\Psi_{1,in}\rangle & = & |\psi_{1}\rangle_{1}  |\psi_{2}\rangle_{2}|\psi_{1}\rangle_{3} \nonumber \\
|\Psi_{2,in}\rangle & = & |\psi_{1}\rangle_{1} |\psi_{2}\rangle_{2} |\psi_{2}\rangle_{3},
\end{eqnarray}
where $|\psi_i\rangle=\alpha_i|H\rangle+\beta_i|V\rangle$, for $i=1,2$, and $H$, $V$ are their horizontal and perpendicular polarization modes \cite{explanation}.
Since $|\psi_{1}\rangle$ and $|\psi_{2}\rangle$ are linearly independent, there exist different permutation symmetries in these two inputs, and we can therefore
design the POVMs, with the elements $\Pi_1$ and $\Pi_2$ satisfying $\Pi_1|\Psi_{2}^{in}\rangle=\Pi_2|\Psi_{1}^{in}\rangle=0$, to unambiguously discriminate them.
In the Bayesian approach \cite{b-h05}, the optimal POVM elements for the unambiguous discrimination of the above inputs are
\begin{eqnarray}
\Pi_1&=&\frac{2}{3}\left(2-\sqrt{\frac{\eta_2}{\eta_1}}\right)I_1\otimes |\Psi^{as}_{23}\rangle\langle\Psi^{as}_{23}|,\nonumber\\
\Pi_2&=&\frac{2}{3}\left(2-\sqrt{\frac{\eta_1}{\eta_2}}\right)I_2\otimes |\Psi^{as}_{13}\rangle\langle\Psi^{as}_{13}|,
\end{eqnarray}
where $|\Psi^{as}_{ij}\rangle=1/\sqrt{2}\left(|H\rangle_i|V\rangle_j-|V\rangle_i|H\rangle_j\right)$ and $I_i$ the identity operators, if the {\it a priori}
probability $\eta_1$ \cite{explanation1} of the data $|\psi_1\rangle$ is in the range of $1/5\leq \eta_1\leq 4/5$. For $4/5\leq \eta_1\leq 1$, the optimal measurement
reduces to the PVM as
\begin{eqnarray}
\Pi_1&=&I_1\otimes |\Psi^{as}_{23}\rangle\langle\Psi^{as}_{23}|\nonumber\\
\Pi_2&=&0;
\end{eqnarray}
and for $0\leq \eta_1\leq 4/5$, the index $1$ and $2$ of the above PVM operators should be interchanged to get the optimal measurement.
In the minimax approach without the available {\it a priori} information of the input data \cite{zy}, on the other hand, the optimal POVM for discriminating 
the inputs in Eq. (2.1) is given as
\begin{eqnarray}
\Pi_1&=&\frac{2}{3}~I_1\otimes |\Psi^{as}_{23}\rangle\langle\Psi^{as}_{23}|,\nonumber\\
\Pi_2&=&\frac{2}{3}~I_2\otimes |\Psi^{as}_{13}\rangle\langle\Psi^{as}_{13}|.
\end{eqnarray}
In both approaches, there is inconclusive measurement result corresponding to the operator $\Pi_0=I-\Pi_1-\Pi_2$.

The physical fundamentals to realize the optimal measurements on the unknown input signals are the teleportation of unknown quantum states 
\cite{teleport} and the deterministic implementation of a POVM on single photon signals \cite{bhe07-2}. A teleportation protocol is applied to transform the triple photon input signals in Eq. (2.1) to the correponding single photon
signals running on different tracks:
\begin{widetext}
\begin{eqnarray}
|\Psi_{1,in}\rangle &=&  |\psi_{1}\rangle_{1}  |\psi_{2}\rangle_{2}|\psi_{1}\rangle_{3} \nonumber\\
&=& c_1|HHH\rangle+c_2|VHH\rangle+c_3|HHV\rangle+\cdots+c_8|VVV\rangle\nonumber\\
&\rightarrow& c_1|0_A, 0_B, H\rangle+c_2|1_A,0_B, H\rangle+c_3|0_A,0_B,V\rangle+\cdots+c_8|1_A,1_B,V\rangle,
\end{eqnarray}
\end{widetext}
where $c_i$ are unknown coefficients, and $0_A$, $1_A$, $0_B$ and $1_B$ the symbols related to the which-path degree of freedom of the photon. After some of
the input signal's polarization degree of freedom is mapped this way to the which-path degree of freedom, we will obtain a single polarization photon 
running on $4$ different tracks 
(the total dimensionality of the quantum system is $8$) so that any POVM performed on it can be deterministically implemented by linear optics. 

To teleport three input qubits in a combined way, we need to have three entangled pairs, e.g., 
\begin{eqnarray}
|\Phi^{+}_1\rangle&=&\frac{1}{\sqrt{2}}\left(|H\rangle|H\rangle+|V\rangle|V\rangle\right),\nonumber\\
|\Phi^{+}_2\rangle&=&\frac{1}{\sqrt{2}}\left(|H\rangle|0_A\rangle+|V\rangle|1_A\rangle\right),\nonumber\\
|\Phi^{+}_3\rangle&=&\frac{1}{\sqrt{2}}\left(|H\rangle|0_B\rangle+|V\rangle|1_B\rangle\right).
\end{eqnarray}
We here define the four Bell states as follows:
\begin{eqnarray}
|\Phi^{\pm}\rangle&\equiv&\frac{1}{\sqrt{2}}\left(|0\rangle|0\rangle\pm|1\rangle|1\rangle\right),\nonumber\\
|\Psi^{\pm}\rangle&\equiv&\frac{1}{\sqrt{2}}\left(|0\rangle|1\rangle\pm|1\rangle|0\rangle\right),
\end{eqnarray}
where $0$ can be $H$, $0_A$, $0_B$, etc, and $1$ can be $V$, $1_A$, $1_B$, etc.
Two of the states in Eq. (2.6) involve both polarization and which-path degrees of freedom. 
How to effectively produce them is what we will discuss in the next section.

\section{Conversion to Single Photon Signal} \label{section3}

\begin{figure}
\includegraphics[width=86truemm]{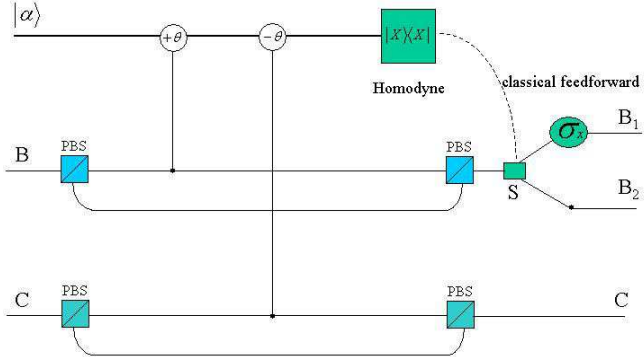}
\caption{(Color online) The quantum non-demolition detection to seperate even parity $\{|H\rangle_B|H\rangle_C, |V\rangle_B|V\rangle_C\}$
from the odd parity $\{|H\rangle_B|V\rangle_C, |V\rangle_B|H\rangle_C\}$ states. Through one of the weak non-linear cross-Kerr, the laser probe initially in
the coherent states $|\alpha\rangle$ will evolve to $|\alpha e^{i\theta}\rangle$ or $|\alpha e^{-i\theta}\rangle$, 
if the single photon is present too.  By a homodyne detection $|\alpha e^{\pm i\theta}\rangle$ resulting from the odd parity can be discriminated from 
$|\alpha\rangle$ from the even. We classically 
feedforward the detection result to the switch, $S$, to direct the odd parity component to the track $B_1$ and the even parity component to the track $B_2$.
On $B_1$ the odd parity is restored to the even by the bit flip and the necessary phase shifters. We use a fictitious {\it controlling action qubit},
$\alpha|0_A\rangle+\beta|1_A\rangle$, acting on the track $B$ photon component to describe the combined action of the CNOT gate and this QND.}
\end{figure}

In this section, we give a rather detailed description of a circuit to map the triple photon input signals to the corresponding single photon signals
as in Eq. (2.5). 

To begin with, we look at the transformation of a double photon polarization state $|\Psi_{in}\rangle=|\psi_1\rangle_1|\psi_2\rangle_2$, where 
$|\psi_i\rangle=\alpha_i|H\rangle+\beta_i|V\rangle$,
to the corresponding single photon state by a combined teleportation procedure. We use two extra photon sources 
$1/\sqrt{2}(|H\rangle_A+|V\rangle_A)$ and $1/\sqrt{2}(|H\rangle_B|H\rangle_C+|V\rangle_B|V\rangle_C)$, and let the first photon on track A control the part B
of the second entangled photon pair through a CNOT gate taking the following action:
\begin{eqnarray}
U_{CNOT}=\left(|H\rangle_A\langle H| \otimes I_B+|V\rangle_A\langle V| \otimes \sigma_{x,B}\right)\otimes I_C.~~~~
\end{eqnarray}
The above equation means that there will be no action on track B if the photon component on track A is in the state of $|H\rangle$ and there will be a bit flip 
$\sigma_{x,B}$ of $\{H,V\}$ on track B if that photon component is in $|V\rangle$. After the joint unitary map $U_{CNOT}$, the total state on the track A, B and C 
will be as follows (here we neglect the common factors for brevity):
\begin{widetext}
\begin{eqnarray}
&&U_{CNOT}\bigl(|H\rangle_A+|V\rangle_A\bigr)\bigl(|H\rangle_B|H\rangle_C+|V\rangle_B|V\rangle_C\bigr)\nonumber\\
&=&|H\rangle_A|H\rangle_B|H\rangle_C+|H\rangle_A|V\rangle_B|V\rangle_C
+|V\rangle_A|V\rangle_B|H\rangle_C+|V\rangle_A|H\rangle_B|V\rangle_C.
\end{eqnarray}
\end{widetext}
This type of non-destructive optical CNOT gate has been experimentally demonstrated \cite{z-z} and, theoretically, a CNOT gate for two independent
single photon polarization qubits can be realized near deterministically \cite{n-m}.

The output state of the CNOT gate in Eq. (3.2) is divided into two sets for part B and C: the even parity 
$\{|H\rangle_B|H\rangle_C, |V\rangle_B|V\rangle_C\}$
and the odd parity $\{|H\rangle_B|V\rangle_C, |V\rangle_B|H\rangle_C\}$, and they can be discriminated near deterministically by a polarization parity 
quantum non-demolition detection (QND) \cite{n-m}. The setup to realize this QND includes a laser probe initially in a coherent state $|\alpha\rangle_p$ and the 
cross-Kerr nonlinearities, which have a Hamiltonian ${\it H}=\hbar\chi a^{\dagger}_s a_s a^{\dagger}_p a_p$, where the signal (probe) mode has the creation and 
destruction operators
given by $a^{\dagger}_s, a_s (a^{\dagger}_p, a_p)$ and $\chi$ the strength of nonlinearity which can be weak. Through the interaction in the weak nonlinear
cross-Kerr in Fig. 1, the photon states on track B, C and the probe evolve together to 
$|\psi_T\rangle\sim (|H\rangle_B|H\rangle_C+|V\rangle_B|V\rangle_C)|\alpha\rangle_p+|H\rangle_B|V\rangle_C
|\alpha e^{i\theta}\rangle_p+|V\rangle_B|H\rangle_C|\alpha e^{-i\theta}\rangle_p$, where $\theta=\chi t$ with $t$ being the interaction time.
$|\alpha\rangle$ and $|\alpha e^{\pm i\theta}\rangle$ can be distinguished between each other by a homodyne-heterodyne measurement on the probe.

Then we classically feedforward two actions to the output on track B: if the the detection result is $|\alpha e^{\pm i\theta}\rangle$, we let the photon component on 
track B go to path $B_1$ and denote the action $0_A$ (the index A means the control from track A photon component); and if the result is 
$|\alpha\rangle$, the photon component on track B is redirected to track $B_2$ with an action named $1_A$. A bit flip (plus the phase shift redressing 
due to the QND) 
on $B_1$ track is performed to restore the odd parity of the photon components on track B and C to the even parity, i.e., 
$|HV\rangle,|VH\rangle\rightarrow |HH\rangle,|VV\rangle$. As the result, we obtain a state,
\begin{widetext}
\begin{eqnarray}
|\psi^{ent}\rangle &\sim&|H\rangle_A|H\rangle_{B_{1}, 0_A}|H\rangle_C+|H\rangle_A|V\rangle_{B_{1}, 0_A}|V\rangle_C
+|V\rangle_A|H\rangle_{B_{2}, 1_A}|H\rangle_C+|V\rangle_A|V\rangle_{B_{2}, 1_A}|V\rangle_C \nonumber\\
&=&(|H\rangle_A|0_A\rangle+|V\rangle_A|1_A\rangle)(|H\rangle_B|H\rangle_C+|V\rangle_B|V\rangle_C),
\end{eqnarray}
\end{widetext}
which can be factorized into the product of two entangled states. $|0_A\rangle$ and $|1_A\rangle$ are defined as two 
{\it controlling switch actions}
on the track B photon component from the photon component on track A. These two fictitious quantum states describe the effective action of 
track B photon component to go to different paths $B_1$ and $B_2$ under the control: 
\begin{eqnarray}
&&|0_A\rangle|H\rangle_B\equiv|H\rangle_{B_{1}, 0_A},~~ |1_A\rangle|H\rangle_B\equiv|H\rangle_{B_{2}, 1_A},\nonumber\\ 
&&|0_A\rangle|V\rangle_B\equiv|V\rangle_{B_{1}, 0_A},~~ |1_A\rangle|V\rangle_B\equiv|V\rangle_{B_{2}, 1_A}.
\end{eqnarray}

It is feasible, therefore, to use these two entangle states to teleport $|\Psi_{in}\rangle=|\psi_1\rangle_1|\psi_2\rangle_2$ to a single
photon polarization state:
\begin{widetext}
\begin{eqnarray}
&&|\psi_1\rangle_1\bigl(|H\rangle_A|0_A\rangle+|V\rangle_A|1_A\rangle\bigr)|\psi_2\rangle_2\bigl(|H\rangle_B|H\rangle_C+|V\rangle_B|V\rangle_C\bigr)\nonumber\\
&&\sim \bigl(|\Phi^+\rangle_{1,A}|\psi_1\rangle_{A_{Ct}}+|\Psi^+\rangle_{1,A}(\sigma_{x,A_{Ct}}|\psi_1\rangle_{A_{Ct}})
+|\Psi^-\rangle_{1,A}(-i\sigma_{y,A_{Ct}}|\psi_1\rangle_{A_{Ct}})+|\Phi^-\rangle_{1,A}(\sigma_{z,A_{Ct}}|\psi_1\rangle_{A_{Ct}})\bigr)\nonumber\\
&&\times \bigl(|\Phi^+\rangle_{2,C}|\psi_2\rangle_{B}+|\Psi^+\rangle_{2,C}(\sigma_{x,B}|\psi_2\rangle_{B})
+|\Psi^-\rangle_{2,C}(-i\sigma_{y,B}|\psi_2\rangle_B)+|\Phi^-\rangle_{2,C}(\sigma_{z,B}|\psi_2\rangle_{B})\bigr),
\end{eqnarray}
\end{widetext}
where the {\it controlling action qubit} is defined as $|\psi_1\rangle_{A_{Ct}}=\alpha_1|0_A\rangle+\beta_1|1_A\rangle$, and $\sigma_x$, $\sigma_y$ and $\sigma_z$ the
Pauli matrices.
With a final restoring unitary transformation $U_{rest}$ on both the tracks $B_1$, $B_2$ and the polarization modes $H$, $V$ according to the information
fed forwarded after the Bell state analysis on track 1, A and 2, C, respectively, we will obtain a single photon polarization state running on track 
$B_1$ and $B_2$ (the total dimensionality of the outputs is 4 as the inputs):
\begin{widetext}
\begin{eqnarray}
|\Psi_{out}\rangle&=&|\psi_1\rangle_{A_{Ct}}|\psi_2\rangle=\bigl(\alpha_1|0_A\rangle+\beta_1|1_A\rangle\bigr)\bigl(\alpha_2|H\rangle_B+\beta_2|V\rangle_B\bigr)\nonumber\\
&=&\alpha_1\alpha_2|H\rangle_{B_{1}, 0_A}+\alpha_1\beta_2|V\rangle_{B_{1}, 0_A}
+\beta_1\alpha_2|H\rangle_{B_{2}, 1_A}+\beta_1\beta_2|V\rangle_{B_{2}, 1_A}.
\end{eqnarray}
\end{widetext}

\begin{widetext}

\begin{figure}
\includegraphics[width=140truemm]{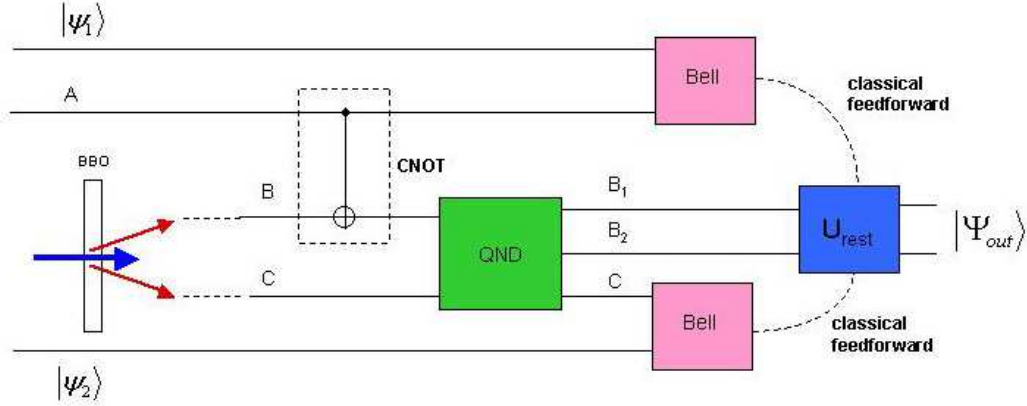}
\caption{(Color online) The scheme to transform a two-photon state, $|\Psi_{in}\rangle=|\psi_1\rangle|\psi_2\rangle$, to a single photon polarization state running on two different tracks, 
$|\Psi_{out}\rangle$. We use a BBO to generate an entangled pair $1/\sqrt{2}(|H\rangle_B|H\rangle_C+|V\rangle_B|V\rangle_C)$ on track B and C, and put a single photon
$1/\sqrt{2}(|H\rangle_A+|V\rangle_A)$ on track A. The CNOT gate can be a modified version of the cited, and the QND unit is shown in Fig. 1.
The results of two Bell state analysis are sent through the classical channels to where the restoration unitary map $U_{rest}$ is performed. $U_{rest}$ is chosen
from a finite set of operations.}
\end{figure}

\end{widetext}

\begin{widetext}

\begin{figure}
\includegraphics[width=120truemm]{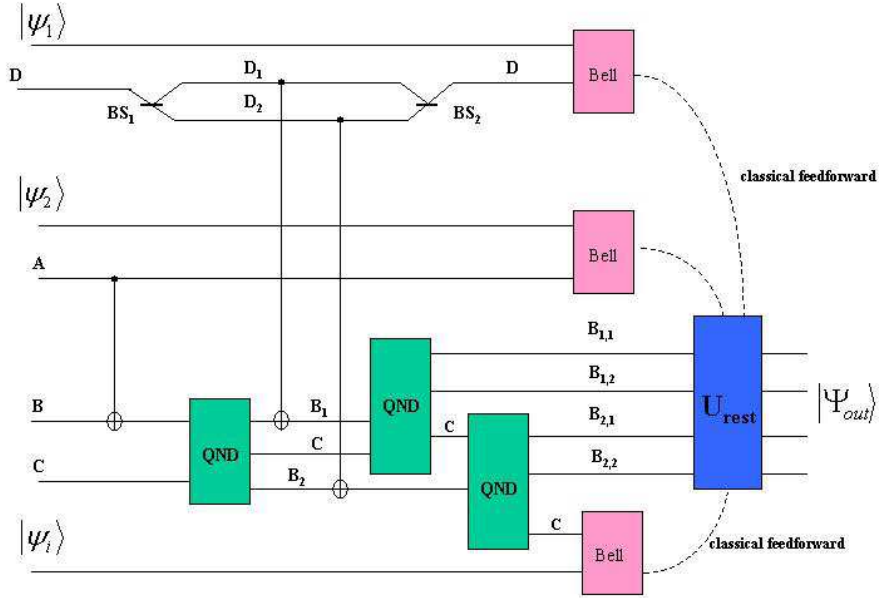}
\caption{(Color online) The scheme to teleport any symmetric unknown three-photon input state $|\psi_{1}\rangle|\psi_{2}\rangle|\psi_{i}\rangle$, for $i=1,2$,
to the corresponding single photon state. One more single photon source is used on track D, and it is split by a 50-50 beam splitter $BS_1$ (and the necessary phase shifters) 
into two paths to control the respective photon components. Another identical beam splitter $BS_2$ merges the photon components on track $D_1$ and $D_2$
to a single path again.}
\end{figure}

\end{widetext}

The complete (and close to deterministic) Bell state analysis setups can be found in \cite{p-a, L-O, s-h}, and also in \cite{teleport1} 
a practical method for teleporting optical qubits is given. 
These experimental schemes can be modified to realize the operations in our setup.
The whole scheme to realize the transformation from double to single photon states is given in Fig. 2.

So far we have put together all necessary ingredients of a circuit to transform the triple photon inputs in Eq. (2.1) to the corresponding single photon
states. To teleport the triple photon inputs to the corresponding single photon states, we just need to continue the above procedure 
before the Bell state analysis by adding one more single photon source, $1/\sqrt{2}(|H\rangle_D+|V\rangle_D)$, and two more CNOT gates and parity QNDs. 
We use a 50-50 beam splitter to split the second single photon into two different paths:
\begin{eqnarray}
|H\rangle_D+|V\rangle_D\rightarrow |H\rangle_{D_1}+|V\rangle_{D_1}+|H\rangle_{D_2}+|V\rangle_{D_2},~~~~
\end{eqnarray}
and the photon components running on $D_1$ ($D_2$) track controls the photon component on $B_1$ ($B_2$) track through an optical CNOT gate. 
Then, performing the polarization parity QNDs on track $B_1$, $C$ and $B_2$, $C$, respectively, and the necessary bit flips and phase redressing, we will 
seperate the photon components on $B_1$ and $B_2$ to those on $4$ tracks $B_{1,1}$, $B_{1,2}$, $B_{2,1}$ and $B_{2,2}$, and thus obtain the output state
which can be factorized into the product of three entangled states (similar to the state in Eq. (3.3)). We are now able to teleport $3$ unknown qubits together
to a single photon polarization state running on $4$ different tracks. The complete scheme of the teleportation of the triple-qubit to a single photon state
is shown in Fig. 3.

\section{Optimal POVM implementation circuit} \label{section4}

The circuit in Fig. 3 generates the following single photon states as the inputs of the POVM modules to discriminate the data $|\psi_1\rangle$ and
$|\psi_2\rangle$:
\begin{widetext}
\begin{eqnarray}
|\Psi^{\prime}_{i,in}\rangle&=&|\Psi_{i,out}\rangle=|\psi_1\rangle_{D_{Ct}}|\psi_2\rangle_{A_{Ct}}|\psi_i\rangle_B\nonumber\\
&=&\bigl(\alpha_1|0_D\rangle+\beta_1|1_D\rangle\bigr)\bigl(\alpha_2|0_A\rangle+\beta_2|1_A\rangle\bigr)\bigl(\alpha_i|H\rangle_B+\beta_i|V\rangle_B\bigr)\nonumber\\
&=&\alpha_1\alpha_2\alpha_i|0_D,0_A,H_B\rangle+\alpha_1\alpha_2\beta_i|0_D,0_A,V_B\rangle+\alpha_1\beta_2\alpha_i|0_D,1_A,H_B\rangle+\alpha_1\beta_2\beta_i|0_D,1_A,V_B\rangle\nonumber\\
&+&\beta_1\alpha_2\alpha_i|1_D,0_A,H_B\rangle+\beta_1\alpha_2\beta_i|1_D,0_A,V_B\rangle+\beta_1\beta_2\alpha_i|1_D,1_A,H_B\rangle+\beta_1\beta_2\beta_i|1_D,1_A,V_B\rangle\nonumber\\
&=&\alpha_1\alpha_2\alpha_i|H\rangle_{B_{1,1}}+\alpha_1\alpha_2\beta_i|V\rangle_{B_{1,1}}+\alpha_1\beta_2\alpha_i|H\rangle_{B_{2,1}}+\alpha_1\beta_2\beta_i|V\rangle_{B_{2,1}}\nonumber\\
&+&\beta_1\alpha_2\alpha_i|H\rangle_{B_{1,2}}+\beta_1\alpha_2\beta_i|V\rangle_{B_{1,2}}+\beta_1\beta_2\alpha_i|H\rangle_{B_{2,2}}+\beta_1\beta_2\beta_i|V\rangle_{B_{2,2}},
\end{eqnarray}
\end{widetext}
where the actions defined in Eq. (3.4) are used in the notation. With $4$ polarization beam splitters (PBS) we can thus obtain the multi-rail single photon signal states running
on $8$ tracks. Any POVM with $n$ elements $\Pi_i$, for $i=1,\cdots,n$, on such single photon states can be deterministically implemented with at most $3n-2$
linear optics modules shown in Fig. 4 \cite{bhe07-2}.

\begin{figure}
\includegraphics[width=60truemm]{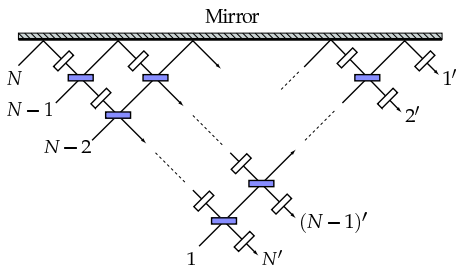}
\caption{(Color online) The unitary transformation module constructed with beam splitters (dark square) and phase shifters
(white square). Any unitary operator represented by an $N\times N$ matrix can be implemented by an $N\times N$ ports module of this kind with the maximum $N(N-1)/2$ of beam splitters \cite{reck}.}
\end{figure}

The POVM for unambiguously discriminating these states satisfies more condition, $\Pi_1|\Psi'_{2,in}\rangle=\Pi_2|\Psi'_{1,in}\rangle=0$, so we can use a 
linear optics scheme, which requires fewer and much simpler modules, 
to realize the unambiguous discrimination of 
$|\Psi'_{1,in}\rangle$ and $|\Psi'_{2,in}\rangle$. 
Such a scheme works with the following unitary map \cite{bhe06-2},
 \begin{eqnarray}
 U_{module}=\left(\begin{array}{cc}(I-\Pi_0)^{\frac{1}{2}}& -\Pi_0^{\frac{1}{2}}\\
 \Pi_0^{\frac{1}{2}} & (I-\Pi_0)^{\frac{1}{2}}\\
\end{array}\right),
 \end{eqnarray}
which is realizable by one $16\times 16$-port module shown in Fig. 4. This module transforms the signals embeded in a larger space of the double of the signal space 
dimension,
 $(|\Psi'_{i,in}\rangle, {\bf 0})^T$, to the components in two orthogonal output subspaces:
 \begin{eqnarray}
 U_{module}\left(\begin{array}{c}|\Psi'_{i,in}\rangle\\
{\bf 0}\end{array}\right)=\left(\begin{array}{c}(I-\Pi_0)^{\frac{1}{2}}|\Psi'_{i,in}\rangle\\
\Pi_0^{\frac{1}{2}}|\Psi'_{i,in}\rangle\end{array}\right).
\end{eqnarray}
The component $(I-\Pi_0)^{\frac{1}{2}}|\Psi'_{i,in}\rangle$ corresponds to the success of the measurement with a probability 
$P=\langle\Psi'_{i,in}|I-\Pi_0|\Psi'_{i,in}\rangle=\langle\Psi'_{i,in}|\Pi_i|\Psi'_{i,in}\rangle$, and the component 
$\Pi_0^{\frac{1}{2}}|\Psi'_{i,in}\rangle$ to the inconclusive result, which occurs with a probability $Q=\langle\Psi'_{i,in}|\Pi_0|\Psi'_{i,in}\rangle$.
It is easy to verify that $|\Psi'_{1,in}\rangle$ and $|\Psi'_{2,in}\rangle$ are mapped to distingushable orthogonal states in the first subspace,
since $\langle\Psi'_{1,in}|I-\Pi_0|\Psi'_{2,in}\rangle=0$.

The linear optics circuit to perform $U_{module}$ can be simplified further. We first decompose the inputs in Eq. (4.1) into
the components in the orthogonal subspaces  
${\cal H}_1=\{|0_D,0_A,V_B\rangle,|0_D,1_A,H_B\rangle,|1_D,0_A,H_B\rangle\}$, 
${\cal H}_2=\{|1_D,0_A,V_B\rangle,|0_D,1_A,V_B\rangle,|1_D,1_A,H_B\rangle\}$
${\cal H}_3=\{|0_D,0_A,H_B\rangle\}$ and ${\cal H}_4=\{|1_D,1_A,V_B\rangle\}$, respectively.
We introduce a unitary map $V_1$,
\begin{widetext}
\begin{eqnarray}
|\Phi_1\rangle&=&\sqrt{\frac{1}{2}}\bigl(|0_D,0_A,V_B\rangle-|0_D,1_A,H_B\rangle\bigr)\nonumber\\
|\Phi_2\rangle&=&\sqrt{\frac{1}{6}}\bigl(|0_D,0_A,V_B\rangle+|0_D,1_A,H_B\rangle\bigr)-\sqrt{\frac{2}{3}}|1_D,0_A,H_B\rangle\nonumber\\                                                                    \nonumber\\
|\Phi_3\rangle&=&\sqrt{\frac{1}{3}}\bigl(|0_D,0_A,V_B\rangle+|0_D,1_A,H_B\rangle+|1_D,0_A,H_B\rangle\bigr),
\end{eqnarray}
in ${\cal H}_1$, and another unitary map $V_2$,
\begin{eqnarray}
|\Phi'_1\rangle&=&\sqrt{\frac{1}{2}}\bigl(|1_D,0_A,V_B\rangle-|1_D,1_A,H_B\rangle\bigr)\nonumber\\
|\Phi'_2\rangle&=&\sqrt{\frac{1}{6}}\bigl(|1_D,0_A,V_B\rangle+|1_D,1_A,H_B\rangle\bigr)-\sqrt{\frac{2}{3}}|1_D,0_A,V_B\rangle\nonumber\\                                                                    \nonumber\\
|\Phi'_3\rangle&=&\sqrt{\frac{1}{3}}\bigl(|1_D,0_A,V_B\rangle+|1_D,1_A,H_B\rangle+|0_D,1_A,V_B\rangle\bigr) 
\end{eqnarray}
\end{widetext}
in ${\cal H}_2$, respectively. These two unitary maps on the modes of the relevant tracks correspond to two simple and identical $3\times 3$-port modules processing the relevant components of the input signals.
In the subspaces ${\cal H}^{\prime}_1=\{|\Phi_1\rangle, |\Phi_2\rangle\}$ and ${\cal H}^{\prime}_2=\{|\Phi'_1\rangle, |\Phi'_2\rangle\}$, $\Pi_0$ of the optimal
POVM is given as a $2\times 2$ matrix \cite{b-h05}:
\begin{eqnarray}
\Pi_0^{(i)}=\left(\begin{array}{cc}-\frac{2}{3}(1-\sqrt{\frac{\eta_1}{\eta_2}}-\sqrt{\frac{\eta_2}{\eta_1}})& -\frac{\sqrt{3}}{6}(2-\sqrt{\frac{\eta_1}{\eta_2}})\\
 -\frac{\sqrt{3}}{6}(2-\sqrt{\frac{\eta_1}{\eta_2}}) & \frac{2}{3}\sqrt{\frac{\eta_1}{\eta_2}}\\
\end{array}\right),~~~~
\end{eqnarray}
for $i=1,2$, and the total $\Pi_0$ is the direct sum,
\begin{eqnarray}
\Pi_0=\Pi_0^{(1)}\oplus\Pi_0^{(2)}\oplus I.
\end{eqnarray}
Therefore, the signal components outside ${\cal H}^{\prime}_1$ and ${\cal H}^{\prime}_2$ only contribute to the inconclusive results, and we only need to process the signal components
in these two subspaces.

\begin{figure}
\includegraphics[width=96truemm]{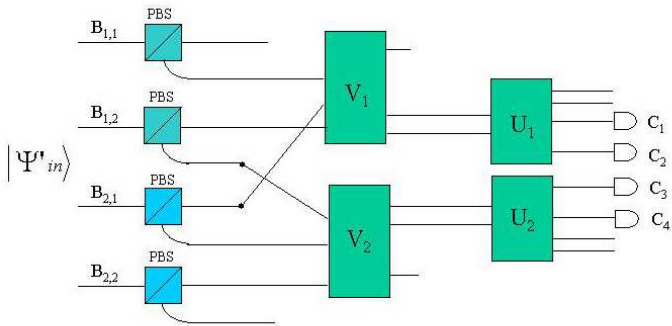}
\caption{(Color online) The layout of the optimal POVM implementation. The input $|\Psi^{\prime}_{in}\rangle$ is the output $|\Psi_{out}\rangle$ in Fig. 3. The polarization beam splitters
(PBS) seperate the input into the $8$ dimensional single photon state in multi-rail representation. The $V_i$ are $3\times 3$-port modules, and 
$U_i$ $4\times 4$-port modules with two input ports black. Only $4$ photon detectors are required to place on the terminals where the useful information is output.
At all other terminals ($2$ of the PBS, $2$ of $V_i$ modules and $4$ of $U_i$ modules), there are only useless signals contributing to the inconclusiveness.
We can set the output in such a pattern: if the detector $C_1$ or $C_3$ clicks, we conclude
that the unknown input is $|\psi_1\rangle$; and if $C_2$ or $C_4$ clicks, the input date must be $|\psi_2\rangle$. From each counter, we detect the photon with
a probability $1/12$, if the data $|\psi_1\rangle$ and $|\psi_2\rangle$ occur with the same {\it a priori} probability.}
\end{figure}

For the {\it a priori} probabilities $\eta_1=\eta_2=1/2$, i.e., the most difficult situation to discriminate the input data, as well as the optimal POVM 
obtained in the minimax approach \cite{zy}, the operator $\Pi_0^{(1)}$ and $\Pi_0^{(2)}$ are therefore reduced to
\begin{eqnarray}
\Pi_0^{(1)}=\Pi_0^{(2)}=\left(\begin{array}{cc}\frac{2}{3}& -\frac{\sqrt{3}}{6}\\
 -\frac{\sqrt{3}}{6}& \frac{2}{3}\\
\end{array}\right).
\end{eqnarray}
The unitary transformations in Eq. (4.2) constructed by these two non-unitary operators are, respectively, implemented by two identical circuits consisting 
of only $3$ beam splitters, one of which performs the unitary map,
\begin{eqnarray}
U_{i,1}=\left(\begin{array}{cc}-\frac{1}{\sqrt{2}}& \frac{1}{\sqrt{2}}\\
 \frac{1}{\sqrt{2}}& \frac{1}{\sqrt{2}}\\
\end{array}\right),
\end{eqnarray}
which diagonalizes $\Pi_0^{(1)}$ and $\Pi_0^{(2)}$, and the other two of which perform the unitary map in the extended space,
\begin{widetext}
\begin{eqnarray}
U_{i,2}&=&\left(\begin{array}{cccc}\sqrt{\frac{1}{3}-\frac{\sqrt{3}}{6}}& & -\sqrt{\frac{2}{3}+\frac{\sqrt{3}}{6}}& \\
 & \sqrt{\frac{1}{3}+\frac{\sqrt{3}}{6}}& & -\sqrt{\frac{2}{3}-\frac{\sqrt{3}}{6}}\\
 \sqrt{\frac{2}{3}+\frac{\sqrt{3}}{6}}& & \sqrt{\frac{1}{3}-\frac{\sqrt{3}}{6}}&\\
 & \sqrt{\frac{2}{3}-\frac{\sqrt{3}}{6}}& &\sqrt{\frac{1}{3}+\frac{\sqrt{3}}{6}} \\
\end{array}\right).\nonumber\\
&&
\end{eqnarray}
\end{widetext}
Through all these unitary transformations the different input signal components from $|\Psi'_{1,in}\rangle$ and $|\Psi'_{2,in}\rangle$, respectively, 
have been mapped to the orthogonal states. Adding one more 50-50 beam splitter to seperate the signals from the different inputs, we realize the target to discriminate the data input: 
the input data is determined to be $|\psi_1\rangle$ if the output photon is recorded by one single photon detector and $|\psi_2\rangle$ if the outout photonis recorded by 
the other. In Fig. 5 we use $U_i$ to represent a whole package of these unitary maps including $U_{i,1}$ and $U_{i,2}$.

By a straightforward calculation considering Eq. (4.4)-(4.5), (4.9) and (4.10), we see that in one of the photon detectors in Fig. 5 there is an average probability 
of $1/12$ to detect the photon. With the detection of the photon, we can draw the conclusion about which the input data is. 
Adding together the contribution of another photon detector placed at one terminal of the other $U_i$ module, we obtain the optimal average success
probability $1/6$ in the discrimnation of a pair of unknown qubits \cite{b-h05}. The layout of the optimal POVM implementation is given in Fig. 5.

\section{Summary and Discussion} \label{section5}
We have constructed a setup to unambiguously and optimally distinguish between any pair of linearly independent single photon polarization qubits, which are completely
unknown to us. This proposed experimental setup applies the state-of-the-art technologies to process quantum information, e.g., the non-destructive optical CNOT gates, 
the complete and near deterministic Bell states analysis including the teleportation of optical qubit, etc., which have been experimentally realized 
\cite{z-z,p-a,L-O, s-h,teleport1}. 
Neglecting the possible error in the operation, the optimal discrimination of any pair of unknown optical qubits can be deterministically performed 
with this device. 
The future advance in these techonologies will make such a device more reliable and less costly.

In principle, we can also generalize the setup to a device that works with multiple reference or data copies, which will increase the success probability
in discriminating the unknown data \cite{bhe06-1}, or a device that discriminates more than two unknowm states \cite{zy}. 
This can be done by teleporting more {\it controlling switch action states}, but the restriction is the cost of the above mentioned experimental facilities.
Another generalization is a setup that unambiguously and optimally discriminates any pair of unknown d-level quantum systems, i.e., {\it qudits}. 
Due to the lack of knowledge about the quantum circuits of the {\it qudits}, it seems that so far there has been no available 
experimental facility to directly discriminate two unknown {\it qudits} unless we first transform them, together with the reference copies, to qubits with a certain possibility. 
These interesting problems, as well as the experimental implementation of the setup we propose in this paper, are worth further study.

\begin{acknowledgments}
The authors acknowledge the support of grants from PSC-CUNY.
\end{acknowledgments}

\bibliographystyle{unsrt}

\begin{thebibliography}{99}
\bibitem{explanation} We can as well use the slightly modified devices to deal with the unknown qubits in other forms of basis, e.g., 
the circular polarization states $|\psi_i\rangle=\alpha_i|L\rangle+\beta_i|R\rangle$, where $|L\rangle=(|H\rangle+i|V\rangle)/\sqrt{2}$,
$|R\rangle=(|H\rangle-i|V\rangle)/\sqrt{2}$.
\end{thebibliography}

\end{document}